\documentclass[12pt,preprint]{aastex}

\newcommand{\be}{\begin{equation}}
\newcommand{\ee}{\end{equation}}

\newcommand{\simless}{\lower.5ex\hbox{$\; \buildrel < \over \sim\;$}}
\newcommand{\simgreat}{\lower.5ex\hbox{$\; \buildrel > \over \sim\;$}} 
\def\lta{\,\raise 0.3 ex\hbox{$ < $}\kern -0.75 em
 \lower 0.7 ex\hbox{$\sim$}\,}
\def\gta{\,\raise 0.3 ex\hbox{$ > $}\kern -0.75 em
 \lower 0.7 ex\hbox{$\sim$}\,} 
\begin{document}

\title{Enhanced Cosmic Ray Flux and Ionization for Star Formation \\ 
in Molecular Clouds Interacting with Supernova Remnants} 

\author{Marco Fatuzzo,$^1$ Fred C. Adams,$^{2,3}$ and Fulvio Melia$^4$} 
 
\affil{$^1$Physics Department, Xavier University, Cincinnati, OH 45207} 

\affil{$^2$Michigan Center for Theoretical Physics, University of Michigan \\
Physics Department, Ann Arbor, MI 48109}  

\affil{$^3$Astronomy Department, University of Michigan, Ann Arbor, MI 48109}

\affil{$^4$Physics Department and Steward Observatory, The University of Arizona, AZ 85721} 

\email{fatuzzo@xavier.edu, fca@umich.edu, melia@physics.arizona.edu}

\begin{abstract} 

Molecular clouds interacting with supernova remnants may be subject to
a greatly enhanced irradiation by cosmic rays produced at the shocked
interface between the ejecta and the molecular gas. Over the past
decade, broad-band observations have provided important clues about
these relativistic particles and indicate that they may dominate over
the locally observed cosmic-ray population by a significant amount. In
this paper, we estimate the enhancement and find that the cosmic ray
energy density can be up to $\sim$1000 times larger in the molecular
cloud than in the field. This enhancement can last for a few Myr and
leads to a corresponding increase in the ionization fraction, which
has important consequences for star formation. Ionization fractions in
molecular cloud cores determine, in part, the rate of ambipolar
diffusion, an important process in core formation and pre-collapse
evolution.  Ionization fractions in newly formed circumstellar disks
affect the magneto-rotational instability mechanism, which in turn
affects the rate of disk accretion. As estimated here, the increased
ionization acts to increase the ambipolar diffusion time by a factor
of $\sim30$ and thereby suppresses star formation. In contrast, the
increased ionization fraction reduces the sizes of dead zones in
accretion disks (by up to an order of magnitude) and thus increases
disk accretion rates (by a comparable factor).

\end{abstract}


\section{INTRODUCTION} 

Although a robust paradigm for star formation within giant molecular
cloud complexes now exists (Shu et al. 1987), the interplay between
stars and the environment from which they are born remains poorly
understood.  This letter considers the interaction between supernova
remnants (SNRs) and molecular clouds (MCs), and studies its effect on
the level of cosmic ray fluxes within star forming regions. The cosmic
ray flux affects the ionization rate, which influences the formation
of molecular cloud cores and the evolution of circumstellar disks.
Specifically, we explore how the cosmic ray production at the SNR/MC
boundary influences the ionization fraction of the molecular gas. This
ionization fraction controls the ambipolar diffusion rate (and hence
the rate of core formation) and the action of magneto-rotational
instabilities (MRI) in circumstellar disks (and hence the rate of disk
accretion).

A growing amount of data concerning the interactions of SNRs and MCs
has been assembled.  Huang \& Thaddeus (1986) found that about half of
SNRs in the outer galaxy are spatially coincident with large molecular
cloud complexes.  More recently, 1720 MHz OH maser emission, a
signpost for SNR-MC interactions (Wardle \& Yusef-Zadeh 2002), has
been observed from nineteen galactic SNRs -- roughly 10\% of the
sample (Yusef-Zadeh et al. 2003). The appearance of this maser
emission is important in that it helps constrain the densities and
temperatures of post-shock gas in these environments. In addition,
magnetic field strengths and orientations can be deduced from Zeeman
splitting and polarization studies (Claussen et al. 1997; Brogan \&
Troland 2001).  High resolution X-ray and CO data have added to the
list of interacting SNR/MC sources (Butt et al. 2001; Dubner et
al. 2004; Byun et al. 2006; Sasaki et al. 2006).

Shocks produced by the interaction of SNRs and MCs can transfer $\sim$
10\% of the SN explosion energy into cosmic-rays (Dorfi 1999, 2000),
and the interaction of these relativistic particles with the ambient
medium can lead to the production of pions and, subsequently, neutral
pion-decay gamma-rays (Aharonian et al. 1994).  Evidence for this
hadronic process has been found through the association of at least
ten EGRET sources with SNRs expanding into MCs (Esposito et al. 1996;
Combi et al. 1999, 2001; Torres et al. 2003); five of these sources
also belong to the aforementioned set of maser SNRs.  While the EGRET
SNRs have gamma-ray luminosities in the 30 MeV -- 10 GeV band ranging
from $L_\gamma \sim 10^{34}$ to $4 \times 10^{36}$ ergs s$^{-1}$, they
have not been detected at TeV energies (Rowell et al. 2000; Buckley et
al. 1997). This finding sets important constraints on the distribution
of energetic particles in these systems and corresponding constraints
on particle-acceleration mechanisms in these high-density environments
(Baring et al. 1999).  These SNRs are also typically radio loud ($F
\simgreat 10$ Jy at 1 GHz), perhaps owing in part to synchrotron
emission from secondary leptons produced via the decay of charged
pions (Fatuzzo \& Melia 2003).

Particle acceleration in the SNR/MC environments can significantly
enhance the cosmic-ray (CR) density above that of the local background
``sea'' surrounding these regions (Aharonian \& Atoyan 1996, hereafter
AA96; Torres et al. 2003).  These energetic particles can increase the
ionization fractions within the molecular cloud, and thereby influence
star formation. One important process in the formation of molecular
cloud cores (stellar birth sites) is the removal of magnetic flux
through ambipolar diffusion. The diffusion coefficient, and the rate,
depends on the ionization fraction of the cloud material. As another
example, an important mechanism for producing accretion in
circumstellar disks is the magneto-rotational instability (MRI) which
depends on the ionization fraction in the disk. The enhanced cosmic
ray flux considered here produces a corresponding increase in the
ionization fraction, which in turn affects the rates of these star
formation processes.

\section{COSMIC-RAY ENHANCEMENT IN SNR/MC ENVIRONMENTS} 

This section estimates the enhancement of cosmic rays in SNR/MC
environments (see also AA96). The EGRET SNRs have inferred gamma-ray
luminosities spanning the range $L_\gamma\sim 10^{34}$ to $4\times
10^{36}$ ergs s$^{-1}$.  This large range in luminosities may be due
in part to the age distributions of the EGRET SNRs. If produced via a
pion-decay mechanism, the gamma-ray luminosity initially increases as
relativistic particles are continually injected into the SNR/MC
environment and their population builds up. The luminosity reaches a
peak once the acceleration mechanism is quenched (Sturner et al.
1996).  Indeed, if 10\% of the SN explosion energy is transferred to
relativistic particles (Dorfi 1991, 2000), the gamma-ray luminosity of
SNR/MC environments would peak at the value
\be
L_\gamma \sim \eta\, \left(\sigma_{pp} n_p c\right) E_{CR}
\approx 2.4 \times 10^{36} \, \hbox{\rm ergs}\, {\rm s}^{-1}\,
\left({n_{H_2}\over 100 \, {\rm cm}^{-3}}\right)\,
\left({\eta\over 0.1}\right)\,
\left({E_{CR}\over 10^{50}\,{\rm ergs}}\right)\;,
\ee
where $\sigma_{pp} = 40$ mb is the $pp$ scattering cross-section,
$\eta$ is the fraction of a relativistic proton's energy that goes
into the $\pi_0$ decay photon channel, $E_{CR}$ is the energy content
of the relativistic particles, and $n_p$ and $n_{H_2}$ are,
respectively, the proton and molecular hydrogen number densities
(i.e., $n_p = 2 n_{H_2}$).\footnote{Molecular clouds are highly
nonuniform, with densities on different scales ranging over several
orders of magnitude.  However, the mean molecular hydrogen density of
MCs is observed to be $\sim 10^2$ cm$^{-2}$} The value of $\eta$ can
be estimated using the approximation for the gamma-ray luminosity due
to the decay of neutral pions
\be
L_\gamma \approx {2 \, \Lambda^0 (\alpha)\over \alpha} \,(\sigma_{pp} n_p c)
 \int E_\gamma {dN_{CR}(E_\gamma)\over dE}\, dE_\gamma\,
\approx {2 \,\Lambda^0 (\alpha) \over\alpha}\, ( \sigma_{pp} n_p c )\, E_{CR}\,,
\ee
where $\alpha$ is the spectral index of the distribution $dN_{CR}/dE$
of relativistic protons, and $\Lambda^0$ is a dimensionless parameter
that depends on $\alpha$ (Crocker et al. 2005).  The above equations
imply $\eta = 2\Lambda^0 / \alpha$, so that $\eta$ = 0.18, 0.10, and
0.60 for $\alpha$ = 2.0, 2.2, and 2.4, respectively. The populations
of relativistic protons injected into SNR/MC environments are expected
to have power-law distributions with $\alpha$ = 2 -- 2.4 (Jones \&
Ellison 1991), in good agreement with the EGRET observations. Since
the diffusion rate depends on the energy of the particles, the
spectrum of relativistic protons may flatten inside a MC so that
$\Delta \alpha \approx 0.2$ (Ballantyne et al. 2006).  Since the
ionization rate depends on the number density of particles (rather
than the energy density), this change in spectral index implies a
corresponding change in number density (by a factor of $\sim2$), but
the exact size of the change depends on the details of the diffusion
process. In future work, a careful assessment of this effect should be
carried out.

Next we assume that particles build up during an injection phase to
their peak values on a timescale of $\sim 10^4$ years, consistent with
estimates of the EGRET SNR ages.  A fraction of these particles will
then diffuse into the cloud with which the SNR is interacting (where
the fraction is determined by the solid angle subtended by the cloud).
Diffusion continues until the cosmic rays undergo $pp$ scattering with
the ambient medium, and thereby experience catastrophic losses, on a
time scale given by
\be
\tau_{pp}\equiv \left(\kappa \sigma_{pp} n_p c\right)^{-1} 
\approx 3 \times 10^5 \,{\rm yrs}\, 
\left({n_{H_2}\over 10^2\,{\rm cm}^{-3}}\right)^{-1}\, , 
\ee
where $\kappa \approx 0.45$ accounts for the inelasticity of $pp$
interactions. Since protons only lose a fraction of their energy in a
given scattering event, the total time of the cosmic ray enhancement
could be larger than $\tau_{pp}$ by a factor of 2 or 3. Setting the
time scale $\tau_{pp}$ equal to the diffusion time scale $\tau_{dif}
\equiv R^2 / D(E)$, where $D(E)$ is the energy-dependent diffusion
coefficient, one can estimate the distance $R(E)$ that particles will
diffuse into the cloud before scattering and losing their energy.
Adopting the expression $D(E) = D_{10}\, (E / 10$ GeV$)^{1/2}$, where
$D_{10} = 10^{28}$ cm$^2$ s$^{-1}$ (AA96), we find
\be 
R(E) = \sqrt{\tau_{pp}\, D(E)} \approx 100 {\rm pc} \, 
\left({n_{H_2}\over 10^2\, {\rm cm}^{-3}}\right)^{-1/2}\,
\left({D_{10}\over 10^{28} \,{\rm cm}^2 \, {\rm s}^{-1}}\right)^{1/2}\,
\left({E\over 10 \,{\rm GeV}}\right)^{1/2}\,.
\ee
As noted in AA96, however, values of $D_{10}\sim 10^{26}$ cm$^2$
s$^{-1}$ may be more applicable in the molecular cloud environment
(Ormes et al. 1988), leading to an estimate of $R$(10 GeV) = 10 pc.
This length scale is comparable to the typical sizes of molecular
clouds.

The energy loss rate for a cosmic-ray due to ionization is weakly
dependent on energy for relativistic protons, and is given by $dE/dt
\sim 3.6 \times 10^{-7}\, n_{H_2}$ eV/s (Mannheim \& Schlickeiser
1994). The corresponding cooling time thus takes the form
\be
\tau_{ci} \approx 8.8 \times 10^6\, {\rm yrs}\,
\left({E\over 10\, {\rm GeV}}\right)\,
\left({n_{H_2}\over 10^2 \,{\rm cm}^{-3}}\right)^{-1}\,,
\ee
which is considerably longer than the $pp$ collision time for all but
the least energetic protons.  The resulting energy density $u_{cr}$
within the MC during the SNR/MC interaction will be
$
u_{cr} \sim 500 \,{\rm eV}\, {\rm cm}^{-3}\,
({E_{CR} / 10^{50} \,{\rm ergs}})\,
({R / 10 \hbox{\rm pc}})^{-3}\,.
$
This value is about $10^3$ times larger than the energy density of the
local cosmic-rays (where $u_{cr} \sim 0.5$ eV cm$^{-3}$). Thus, a
significant enhancement of cosmic ray energy density can be achieved in
molecular clouds through their interactions with SNRs.

\section{ENHANCEMENT OF THE IONIZATION FRACTION} 

The number density of ionized particles in a MC environment depends on
the ionization rate (due to cosmic-rays) and a complex recombination
process. For steady-state conditions, the ionization fraction $x
\equiv n_e / n_{H_2}$ is well approximated (Elmegreen 1979; Shu 1983,
1992) by the expression
\be
x\sim 10^{-5}\,
\left({\zeta / 10^{-17}\,{\rm s}^{-1}}\right)^{1/2}\,
\left({n /  10^2 {\rm cm}^{-3}}\right)^{-1/2} \; . 
\ee
This expression is valid for $n \simless 10^6$ cm$^{-3}$ $(\zeta /
10^{-17}$ s$^{-1}$). Assuming that $\zeta$ scales linearly with the
cosmic ray energy density, i.e., $\zeta = 10^{-17} \,{\rm s}^{-1}\,
({u_{cr} / 0.5\, {\rm eV}\, {\rm cm}^{-3}})$, the ionization fraction
in molecular clouds interacting with SNRs can be enhanced by up to a
factor of $\sim 30$. This scaling is strictly valid when $n \simless
10^9$ cm$^{-3}$ for the enhanced cosmic ray flux considered here. The
process of ambipolar diffusion occurs primarily at lower densities,
and will be fully affected by this enhancement. In circumstellar
disks, however, the densities are larger near the disk midplane, and
the full ionization enhancement might not be realized. Notice that the
ionization rate depends on the number density of cosmic rays, rather
than the energy density. If the spectrum $dN_{CR}/dE$ is invariant,
the two quantities scale together. However, the index $\alpha$ of the
spectrum could vary (\S 2), so that the scaling could differ by a
factor of $\sim4$ (which affects the ionization fraction by a factor
of 2).

The timescale $\tau_i$ required to reach this enhancement level can be
estimated by setting the ionization rate $\dot n_e = \zeta n$ equal to
the ratio $n_e / \tau_i$. Using the expression for the ionization
fraction $x$ given by equation (6), this timescale estimate becomes
$\tau_i$ = $x / \zeta \approx 3 \times 10^4$ yr $(\zeta /
10^{-17}\,{\rm s}^{-1})^{-1/2}$ $(n / 10^2\,{\rm cm}^{-3})^{-1/2}$,
only about 1000 yr when $\zeta \sim 10^{-14}$ s$^{-1}$. The
ionization enhancement is thus established at virtually the same 
time as the CR enhancement.  Of course, this result also means that
the ionization enhancement will decay on a comparably short timescale,
as soon as the CR enhancement is removed due to $pp$ scattering (up to
1 Myr, as estimated in \S 2; this time can be extended, as discussed
below).

The discussion so far considers the MC environment to be uniform.
However, these MC regions are highly nonuniform, exhibiting
hierarchical structure that can be characterized in terms of dense
clumps and cores surrounded by an interclump gas with $n \sim 5 - 25$
cm$^{-3}$.  Clumps have characteristic densities $n \sim 10^3$
cm$^{-3}$ and radii $r \approx 0.1 - 1$ pc.  These dense regions
occupy a relatively small fraction (2 -- 8 \%) of the cloud volume,
but can account for most of its mass (Williams et al. 1995).  As a
result, cosmic rays propagating through the MC environment spend most
of their time in a lower density medium ($n_{H_2} \sim 10$ cm$^{-3}$),
so the $pp$ collision timescale would be $\sim 3$ Myr. Of course, this
point is somewhat diminished by the fact that these particles also
spend a small fraction of their life ($\sim 2 - 8$\%) in the denser
regions. If these particles can survive without undergoing $pp$
scattering for $\sim 1$ Myr, the enhancement in cosmic ray energy
density can last for a few Myr.

The interaction of relativistic protons with the MC medium leads to
the production of both neutral and charged pions, with each species
produced in equal numbers.  As a result, $pp$ scattering transfers
roughly 3\% of the initial cosmic-ray energy to secondary electrons
and positrons -- the byproducts of charged pion decays.  The injection
of these leptons can increase the energy density of cosmic-rays above
the field by as much as a factor of $\sim 30$, and increases the
ionization fraction by a factor of $\sim 5$ (assuming that $\zeta
\propto u_{cr}$).  The peak of the injected secondary lepton
distribution occurs at $\sim 60$ MeV.  Near this peak energy, leptons
traversing neutral atomic hydrogen cool predominantly via electronic
excitation on a timescale $\tau_c \sim 7 \times 10^4 (n_p /10^2$
cm$^{-3})^{-1}$ yr (Gould 1975).  The enhancement in the ionization
fraction due to secondary leptons would therefore only last for a
comparably short time.

\section{EFFECTS ON STAR FORMATION} 

The enhancement in cosmic ray flux, which increases ionization levels,
can have a significant impact on star formation. Ionization affects
the coupling between magnetic fields and the (largely neutral) gas. On
scales of $\sim 0.1$ pc, increased ionization leads to greater
coupling between gas and the magnetic fields, and acts to slow down
star formation. On smaller scales of $\sim1$ AU, increased ionization
in circumstellar disks allows for MRI to remain active over a greater
extent of the disk and thereby allows for larger disk accretion rates.

In considerations of ambipolar diffusion, the density of ions takes
the form $\rho_i = {\cal C} \rho_n^{1/2}$ because the volumetric rate
of recombinations is proportional to $n_e n_i \sim n_i^2$ and the
volumetric rate of ionization is proportional to $\zeta n_n$ (Shu
1992). As a result, the constant ${\cal C} \propto \sqrt{\zeta}$.
Standard considerations of ambipolar diffusion (Mouschovias 1976; Shu
1983, 1992) show that the effective diffusion constant is given by 
$D \sim v_A^2 / (\gamma {\cal C} \rho) \propto \zeta^{-1/2}$, where
$v_A$ is the Alfv{\'e}n speed and $\gamma$ is the drag coefficient
between ions and neutrals. As the cosmic ray flux increases, the
diffusion coefficient decreases as the square root of the flux. A
1000-fold increase in the cosmic ray flux (\S 2) thus decreases the
effective diffusion constant by a factor of $\sim30$, and increases
the ambipolar diffusion time scale by the same factor. Current
observations indicate that the diffusion process is too slow to
account for the observed statistics of starless molecular cloud cores
(e.g., Jijina et al. 1999), so that some mechanism to speed up the
process is required, even in the absence of supernova enhancement of
the cosmic ray flux (Ciolek \& Basu 2001, Zweibel 2002, Fatuzzo \&
Adams 2002). The factor of $\sim30$ increase in timescale due to
cosmic ray enhancement essentially shuts down the process. Although
supernovae have often been invoked as a means to trigger star
formation (Elmegreen 1998), this mechanism provides a channel for
supernovae to inhibit star formation.

An enhancement in cosmic ray flux also influences disk accretion for
star/disk systems associated with the interaction region.  A growing
consensus in the field holds that disk accretion is produced by an
effective viscosity that is driven by turbulence, which in turn is
driven by MHD instabilities such as MRI (Balbus \& Hawley 1991). In
order for MRI to operate, and hence for disk accretion to take place,
the ionization fraction must be sufficiently high so that the gas is
well coupled to the field. The inner disk can be ionized by collisions
(where the number density and temperature are high), and the outer
disk can be ionized by standard values of the cosmic ray flux, but
intermediate regions may have dead zones where ionization is too low
(Gammie 1996).  The enhanced cosmic ray flux acts to increase the
fraction of the disk that is active, i.e., sufficiently ionized for
MRI to operate.

To illustrate the possible effects of an enhanced cosmic ray flux,
consider the disk models of Gammie (1996), where the cosmic ray flux
has its standard (unenhanced) value.  The flux is assumed to be
ineffective as an ionization source after it falls to 1/e times its
initial value. The attenuation column density for cosmic ray
ionization is $\Sigma_0 \approx 100$ g/cm$^2$ (Umebayashi \& Nakano
1981).  As a result, only the uppermost 100 g/cm$^2$ of the disk (and
a corresponding layer on the other side) experience enough cosmic ray
ionization for MRI to operate. With a cosmic ray flux that is enhanced
by a factor $\cal F$, the disk will experience cosmic ray ionization
down to a larger column density $\Sigma_\ast \approx (1 + \ln {\cal
F}) \Sigma_0$. For an optimal enhancement of ${\cal F} \approx 1000$,
the active column density becomes $\Sigma_\ast \approx 8 \Sigma_0
\approx 800$ g/cm$^2$ ($\sim8$ times more disk material can be
ionized). To put this result in perspective, consider a circumstellar
disk with mass $M_d = 0.05 M_\odot$, radius $r_d$ = 30 AU, and surface
density profile $\Sigma \propto 1/r$.  The total surface density at 1
AU is thus $\Sigma$ (1 AU) $\approx$ 2400 g/cm$^2$.  Since the upper
and lower 800 g/cm$^2$ of the column can be ionized with an enhanced
cosmic ray flux, about 2/3 of the disk can remain active at $r$ = 1
AU. As a result, the dead zone subtends only the central 33\% of the
disk column density, compared with 92\% for no cosmic ray enhancement.
Since both the disk properties and the cosmic ray flux enhancement
will vary from system to system, an extensive exploration of parameter
space is warranted.  Nonetheless, this calculation illustrates that
supernova enhancements of the cosmic ray flux can have a substantial
impact on disk accretion. In the most extreme case, the disk accretion
rate can be an order of magnitude larger than in regions with no cosmic 
ray enhancement.

\section{CONCLUSIONS}  

This letter has shown that a supernova interacting with a molecular
cloud can produce a substantial enhancement in the flux of cosmic rays
within the cloud.  These cosmic rays can diffuse a distance $R \sim
10$ pc into the molecular cloud before undergoing $pp$ scattering
leading to catastrophic energy losses.  The resulting flux enhancement
is estimated to be a factor of $\sim 10^3 (R / 10$ pc)$^{-3}$ within
the diffusion region, and can last for a few Myr after cosmic-ray
acceleration ceases.  The corresponding enhancement in the ionization
fraction within the molecular cloud is estimated to be $\sim 30 (R /
10$ pc)$^{-3/2}$. This cosmic-ray enhancement may have a significant
impact on star formation processes, primarily by affecting the
ionization fraction. The enhanced ionization fraction in MC cores, the
sites of protostellar collapse, increases the ambipolar diffusion time
scale by a factor of $\sim30$, so that supernovae inhibit further star
formation via this mechanism.  The enhanced cosmic ray flux also
increases the ionization levels within circumstellar disks.  In this
setting, more of the disk can actively support MRI, so that supernovae
enhancement of cosmic ray flux leads to increased rates of disk
accretion, by a factor of $\sim$8.

During the past decade, multiwavelength observations have provided
important clues about cosmic ray acceleration resulting from the
interaction between supernova remnants and molecular clouds.  While OH
maser emission has helped constrain the environment of the 
acceleration sites, EGRET observations have helped constrain the
resulting particle distributions.  Additionally, since the gamma-ray
emission above $\sim 100$ MeV is believed to result primarily from the
decay of neutral pions produced by the scattering of the cosmic rays
with the ambient medium, future GLAST observations promise to advance
our understanding of how cosmic rays diffuse from their acceleration
sites and interact with the molecular cloud environment. Along with
these future observations, additional theoretical work should be
carried out to explore the details of the diffusion of cosmic rays
into molecular clouds, the enhanced ionization fractions produced
therein, and the subsequent effects on star formation. 

\bigskip 
{} 
\bigskip 
This work was supported by Xavier University through the Hauck
Foundation.  This work was supported at the University of Michigan by
the Michigan Center for Theoretical Physics, and by NASA through the
Astrophysics Theory Program (NNG04GK56G0) and the Spitzer Space
Telescope Theoretical Research Program.  This work was supported 
at The University of Arizona by NSF (AST-0402502).

\bigskip 
%\newpage 

%\newpage 
%\begin{figure}
%\figurenum{1}
%{\centerline{\epsscale{0.90} \plotone{f1.ps} }}
%\figcaption{For initial states }  
%\end{figure}

\end{document}